\documentclass[11pt]{article}

\def\mytitle#1{\setcounter{equation}{0}
\setcounter{footnote}{0}
\begin{flushleft}\Large\textbf{#1}\end{flushleft}
\vspace{0.25cm}}
\def\myname#1{\leftline{{\large #1}}\vspace{-0.13cm}}
\def\myplace#1#2{\small\begin{flushleft}\textit{#1}\\
\texttt{#2}\end{flushleft}}

\def\myclassification#1{\small\noindent
Pacs no :
       #1\vspace{0.5cm}}
\usepackage{graphicx}
\begin{document}

\mytitle{Interacting Holographic Dark Energy Model as a Dynamical system and the Coincidence Problem}

\myname{Ritabrata Biswas\footnote{biswas.ritabrata@gmail.com}}
\vskip0.2cm \myname{Nairwita
Mazumder\footnote{nairwita15@gmail.com}} \vskip0.2cm
\myname{Subenoy
Chakraborty\footnote{schakraborty@math.jdvu.ac.in}} \vskip0.2cm
\myplace{Department of Mathematics, Jadavpur University,
Kolkata-700 032, India.} { }

\begin{abstract}
We examine the evolution of a holographic cosmological model with
future event horizon as the infrared cut-off and dark matter and
dark energy do not evolve independently $-$ there is interaction
between them. The basic evolution equations are reduced to an
autonomous system and corresponding phase space is analyzed.\\
Keywords : Dynamical System, Phase plane, Holographic Dark energy.
\end{abstract}
\myclassification{04.20.-q, 04.40.-b, 95.35.+d, 98.80.Cq}
\section{Introduction}
Recent observational evidences particularly from Type Ia
supernovae and Cosmic Microwave Background(CMB) speculate the
existence of both gravitating and non-gravitating type of matter.
There is a substantial amount of gravitating matter non-baryonic
in nature and is termed as Dark Matter(DM)
\cite{Dunkley1,Tegmark1,Percival1,Riess1,Perlmutter1}. On the
other hand, the non-gravitatng matter, known as Dark Energy(DE) is
the mysterious agent for the present phase of cosmic accelerated
expansion. It is only known for certain that DE has huge negative
pressure(comparable to its energy density) and there is sufficient
reason to assume an even distribution of it over the space[for
details see ref \cite{Amendola1}]. Although DM energy density is
expected to decrease at a faster rate than the density of DE
throughout the evolution, interestingly they have comparable
magnitude today.. This surprising matching is known as the
'coincident problem'. To resolve this problem, use of tracker
fields \cite{Zlatev1} and oscillating DE models \cite{Nojiri1} are
normally employed. But recently, there arises a third possibility
\cite{Mangano1,He1,Ma1,Pavon1} by introducing DE and DM interact
through an additional coupling term in the fluid equation. In the
present work, we choose the third possibility as a solution of the
problem.

To have some inside about the unknown and mysterious nature of DE,
many people have suggested that DE should be compatible with
Holographic principle ,namely "the number of relevant degrees of
freedom of a system dominated by gravity must vary along with the
area of the surface bounding the system"\cite{Hooft1}. Such a DE
model is known as Holographic DE(HDE) model. Further the energy
density of any given region should be bound by that ascribed to a
Schwarzschild black hole(BH) that fills the same volume
\cite{Cohen1,Li1}. Mathematically, we write $\rho_{D}\leq
M_{p}^{2}L^{-2}$, where $\rho_{D}$ is the DE density, $L$ is the
size of the region(or infrared cut off) and $M_{p}=\left(8\pi
G\right)^{-\frac{1}{2}}$ is the reduced Planck mass. Usually, the
DE density is written as
\begin{equation}\label{1}
\rho_{D}=\frac{3M_{p}^{2}c^{2}}{L^{2}}
\end{equation}
Here the dimensionless parameter '$c^2$' takes care of the
uncertainties of the theory and for mathematical convenience the
factor 3 has been introduced. In HDE paradigm
\cite{Cohen1,Li1,Horava1, Thomas1,Hsu1,Pavon2}one determines an
appropriate quantity to serve as an IR cut off for the theory and
imposes the constraint that the total vacuum energy in the
corresponding maximum value must not be greater than the mass of a
BH of the same size. By saturating the inequality one identifies
the acquired vacuum energy as HDE. Although the choice of the IR
cut off has raised on discussion in the literature \cite{Li1,
Pavon2, Gong1,Guberina1,Setare1,Setare2}, it has been shown, and
it is generally accepted, that the radius of the event horizon of
the universe($R_{E}$) the most suitable choice for the IR cut off
where $R_{E}$ is defined as \cite{Hsu1}:
\begin{equation}\label{2}
L=R_{E}=a\int_{t}^{\infty}\frac{dt}{a}.
\end{equation}
Now for the interacting DM and DE to resolve the coincidence
problem (as mentioned above), the interaction term is chosen in
the present work in the following two ways : (a) usually, the
interaction term is chosen as $A H \rho_{m}+B H \rho_{D}$ where
$\rho_{m}$ and $\rho_{D}$ are DM and DE densities and $A$, $B$ are
dimensionless constants. For convenience we shall choose
$A=B=3b^{2}$,i.e., interaction of the form $3b^{2}H\rho$ where
$\rho(=\rho_{m}+\rho_{D})$ is the total energy density. (b) a
natural and physical variable interaction term is of the form
$\gamma \rho_{m}\rho_{D}$ with $\gamma$ a dimension full
($\frac{L^{3}}{mt}$) constant. Note that this interaction term
vanishes (as expected) if any one of the energy densities is zero
while the interaction term grows with the increase of both the
energy densities. Further, such an interaction term gives the best
fit to observations \cite{Ma1} for HDE models. In the present work
we choose DM in the form of pure dust while the HDE, as perfect
fluid with equation of state $p_{D}=\omega_{D} \rho_{D}$.

\section{Basic Equations}\label{chapter2}
We consider our universe to be homogeneous and isotropic flat FRW
model and assume that it fills with DM in the form of dust (having
energy density $\rho_{m}$) and HDE in the form of a perfect fluid
having equation of state $p_{D}=\omega_{D}\rho_{D}$ where
$\omega_{D}$ is variable.

The Einstein field equations for spatially flat model are
\begin{equation}\label{3}
3H^{2}=\rho_{m}+\rho_{D}
\end{equation}
and
\begin{equation}\label{4}
2\dot{H}=-\rho_{m}-\left(1+\omega_{D}\right)\rho_{D}
\end{equation}
where for simplicity we choose $8\pi G =1=c$

The conservation equations for the fluids are
\begin{equation}\label{5}
\dot{\rho_{m}}+3H\rho_{m}=Q
\end{equation}
and
\begin{equation}\label{6}
\dot{\rho_{D}}+3H\left(1+\omega_{D}\right)\rho_{D}=-Q
\end{equation}
It is to be noted that for $Q>0$, energy is transferred from DE to
DM and opposite is the situation for $Q<0$. As $Q<0$ would worsen
the coincidence problem so we choose $Q>0$ throughout the work.
Further, for validity of second law of thermodynamics and
Lechatelier's principle \cite{Pavon1} one must take $Q>0$. Also it
should be mentioned that baryonic matter is not included in the
interaction due to the constraints imposed by local gravity
measurements \cite{Peebles1,Hagiwara1}.

Using the field equations (\ref{3})and (\ref{4}), the acceleration of the universe is given by
\begin{equation}\label{7}
\ddot{a}=a\left(\dot{H}+H^{2}\right)=-\frac{a}{6}\left\{\rho_{m}+\left(1+\omega_{D}\right)\rho_{D}\right\}
\end{equation}
which shows that for the present accelerating phase it is
necessary (but not sufficient) to have $\omega_{D}<-\frac{1}{3}$.

Using the density parameters
\begin{equation}\label{8}
\Omega_{m}=\frac{\rho_{m}}{3H^{2}}~~~,~~~\Omega_{D}=\frac{\rho_{D}}{3H^{2}}
\end{equation}
the Einstein equation (\ref{3}) can be written as
\begin{equation}\label{9}
\Omega_{m}+\Omega_{D}=1
\end{equation}
Introducing $u=\frac{\rho_{m}}{\rho_{D}}$ as the ratio of the energy densities we have
\begin{equation}\label{10}
\Omega_{m}=\frac{u}{1+u}~~~,~~~\Omega_{D}=\frac{1}{1+u}
\end{equation}
\section{Detailed Calculations for $Q=3b^{2}H\rho$}\label{chapter3}

Using the conservation equations (\ref{5}) and (\ref{6}) and the
energy density of HDE from equation (\ref{1}), we have the
expression for the equation of state parameter as

\begin{equation}\label{11}
\omega_D=-
\frac{1}3-\frac{2\sqrt{\Omega_D}}{3c}-\frac{b^2}{\Omega_D}<-\frac{1}{3}
\end{equation}

which shows that there will be always acceleration.\\

The evolution of the density parameter $\Omega_{D}$ is given by
\begin{equation}\label{12}
\dot{\Omega_{D}}=H\Omega_{D}^{2}\left(1-\Omega_{D}\right)\left[\frac{1}{\Omega_{D}}+\frac{2}{c\sqrt{\Omega_{D}}}-\frac{3b^{2}}{\Omega_{D}\left(1-\Omega_{D}\right)}\right]
\end{equation}
and hence the ratio of the energy densities evolves as
\begin{equation}\label{13}
\dot{u}=H\left[-u\left\{1+\frac{2}{c\sqrt{1+u}}\right\}+3b^{2}\left(1+u\right)\right]
\end{equation}
The Friedmann equation (\ref{4}) and the conservation equation
(\ref{6}) can be converted(after a bit simplification) into an
autonomous system as
\begin{equation}\label{14}
\dot{\rho}_{D}=2\rho_{D}\left[\frac{\sqrt{\rho_{D}}}{\sqrt{3}c}-H\right]
\end{equation}
and
\begin{equation}\label{15}
\dot{H}=\frac{1}{2}\left[3H^{2}\left(1-b^{2}\right)-\frac{\rho_{D}}{3}-\frac{2}{3\sqrt{3}c}\frac{\rho_{D}^{\frac{3}{2}}}{H}\right]
\end{equation}

\begin{figure}
\includegraphics[height=2.6in, width=2.6in]{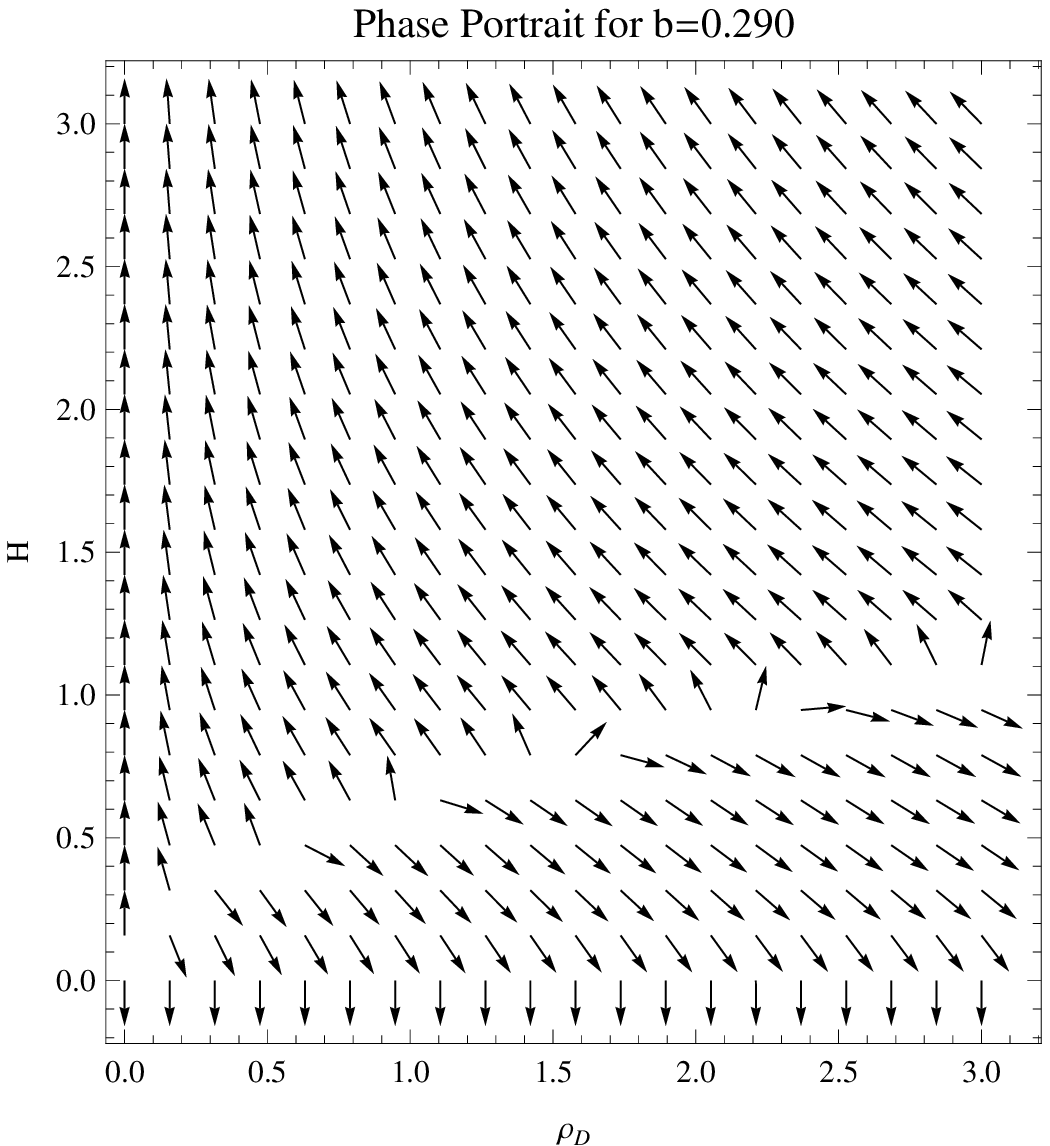}~~
\includegraphics[height=2.6in, width=2.6in]{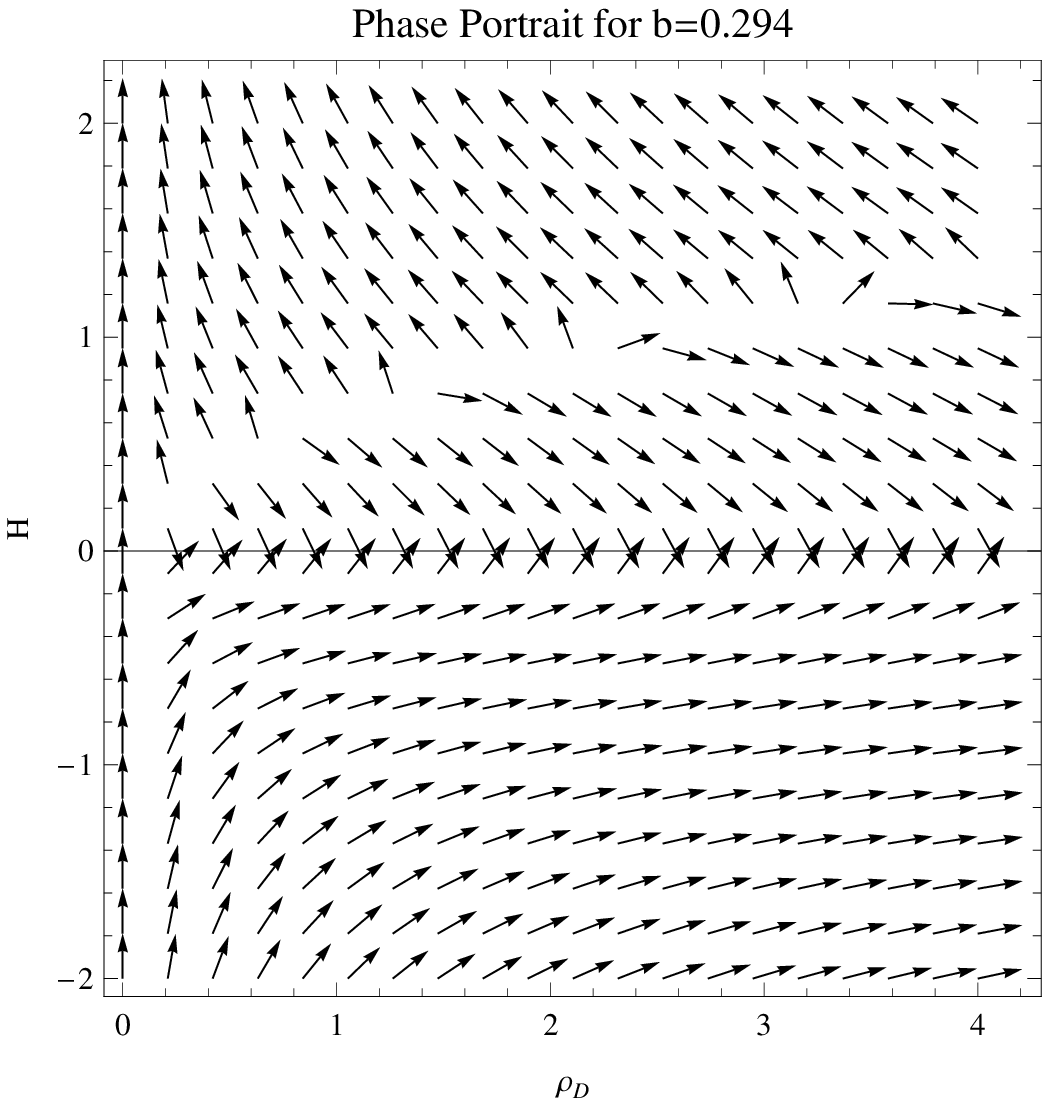}\\
~~~~~~~~~~~~~~~~~~~~~~~~~~~~~~~~~~~~~~~~~~~~~~~~~~~~~~~~~~~~~~~~~Fig.1(a)
.~~~~~~~~~~~~~~~~~~~~~~~~~~~~~~~~~~~~~~~~~~~~~~~~~~~~~~~~~~~~~~~~Fig.1(b)
\\\\

Fig. 1(a)-1(b) represent the variation of $\rho_D - H$. Though in
Fig 1(b) the whole region is not a physically valid region but for
better understanding about the system we have drawn the whole
region.

\hspace{1cm} \vspace{2cm}

\end{figure}

The dynamical system has a line of critical points along the
parabola $\rho_{D}=3c^{2}H^{2}$ in the phase plane
$\left(\rho_{D},~H\right)$ with the restriction $b^{2}=1-c^{2}$.
Then the linearized matrix $A$ has
$trace(A)=H\left(1-4c^{2}\right)$ and $determinant (A)=0$. So the
phase paths form a family of parabolas \cite{Perko1}. The phase
portrait for different choices of the parameter '$b$' are shown in
figures 1(a) and 1(b). Note that along the line of critical points
$\Omega_{D}=c^{2}$ and
$\omega_{D}=-\frac{1}{c^{2}}<-1,~u=\frac{b^{2}}{c^{2}}$. Hence
along the phase paths the ratio of the energy densities bears a
constant
value and the universe will be in the phantom era.\\

Further, fixed points corresponding to $\dot{u}=0$ is essentially
a cubic equation which has at least one real root say, $u_{f}$.
Then the parameter '$b^{2}$' can be estimated by the fixed point
as

\begin{equation}\label{16}
b^{2}=\left(\frac{u_f}{1+u_f}\right)\left(\frac{1}{3}+\frac{2}{3c\sqrt{1+u_f}}\right)
\end{equation}
Now, to analyse the stability of the fixed point we write
\begin{equation}\label{17}
u'=\frac{du}{dx}=\frac{du}{dt}\frac{dt}{dx}=\frac{\dot{u}}{H}=-u\left\{1+\frac{2}{c\sqrt{1+u}}\right\}+3b^{2}\left(1+u\right)
\end{equation}
where $x=\ln a$.

Then at the fixed point
\begin{equation}\label{18}
\frac{du'}{du}=\left.
3b^{2}-1-\frac{(u+2)}{3(1+u)^{\frac{3}{2}}}\right|_{u=u_{f}}
\end{equation}

$$=-\frac{1}{1+u_{f}}-\frac{2-u_{f}}{c\left(1+u_{f}\right)^{\frac{3}{2}}}<0.$$
Hence the fixed point $u_{f}$ is a stable one.

Moreover the conservation equations (\ref{5}) and (\ref{6}) can be written as
\begin{equation}\label{19}
\dot{\rho}_{m}=\sqrt{3\left(\rho_{m}+\rho_{D}\right)}\left[b^{2}\rho_{D}-\left(1-b^{2}\right)\rho_{m}\right]
\end{equation}
\begin{equation}\label{20}
\dot{\rho}_{D}=-\sqrt{3\left(\rho_{m}+\rho_{D}\right)}\left[b^{2}\rho_{m}+\left(1+\omega_{D}+b^{2}\right)\rho_{D}\right]
\end{equation}
From equation(\ref{20}) we see that $\dot{\rho}_{D}\leq 0$, i.e.,
DE density decreases at least in the quintessence era. Also from
the equation (\ref{19}), if we assume \cite{Lip1} $\rho_{D}$ to be
sufficiently large initially then matter density increases in the
early phase and subsequently it decreases with $\dot{\rho}_{m}=0$
along the straight line $\rho_{m}=\frac{b^{2}}{1-b^{2}}\rho_{D}$
in the $(\rho_{m},~ \rho_{D})-$ plane. Then in the phantom era
$\left(\omega_{D}<-1\right)$, $\rho_{D}$ may begin to increase and
dominate over DM.

Thus the present model of the universe shows a DE domination
initially and subsequently the universe evolve with DM domination.
Then there may be DE dominated phase at late time as predicted by
observation. Hence this scenario is favourable for the present
universe.

Further, from the point of view of the coincidence problem we see
that $u\left(=\frac{\rho_{m}}{\rho_{D}}\right)$ is less than unity
in the early phase of the universe and then it gradually
increases. $u\sim o(1)$ before $\dot{\rho}_{m}=0$ or after
$\dot{\rho}_{m}=0$ or along the straight line
$\rho_{m}=\frac{b^{2}}{1-b^{2}}\rho_{D}$ in the
$\left(\rho_{m},~\rho_{D}\right)$-plane provided
$b^{2}>~or~<~or~=\frac{1}{2}$. Though the coincidence problem has
partial solution around the straight line
$\rho_{m}=\frac{b^{2}}{1-b^{2}}\rho_{D}$, but it does not give any
explaination for $u\sim o(1)$ in the present scenario.
\section{Calculation details for $Q=\gamma \rho_{m} \rho_{D}$}

Proceeding exactly as in the previous section the expression for
equation of state parameter and the evolution of the density
parameter are given by
\begin{equation}\label{21}
\omega_{D}=-\frac{1}{3}-\frac{2\sqrt{\Omega_{D}}}{3c}-\gamma H\left(1-\Omega_{D}\right)<-\frac{1}{3}
\end{equation}
and
\begin{equation}\label{22}
\dot{\Omega}_{D}=H\Omega_{D}\left(1-\Omega_{D}\right)\left[1-3\gamma
H \Omega_{D} + \frac{2\sqrt{\Omega_{D}}}{c}\right]
\end{equation}
So the evolution of the ratio of the energy densities is described
as
\begin{equation}\label{23}
\dot{u}=3H u\left[-\frac{1}{3}-\frac{2}{3c\sqrt{1+u}}+\frac{\gamma H}{1+u}\right]
\end{equation}
Also the Friedmann equation (\ref{4}) can be written as
\begin{equation}\label{24}
\dot{H}=-\frac{3H^{2}}{2}\left[-\frac{\Omega_{D}}{3}-2\frac{\Omega_{D}^{\frac{3}{2}}}{3c}+1-\gamma H\Omega_{D}\left(1-\Omega_{D}\right)\right]
\end{equation}

\begin{figure}
\includegraphics[height=2.6in, width=2.6in]{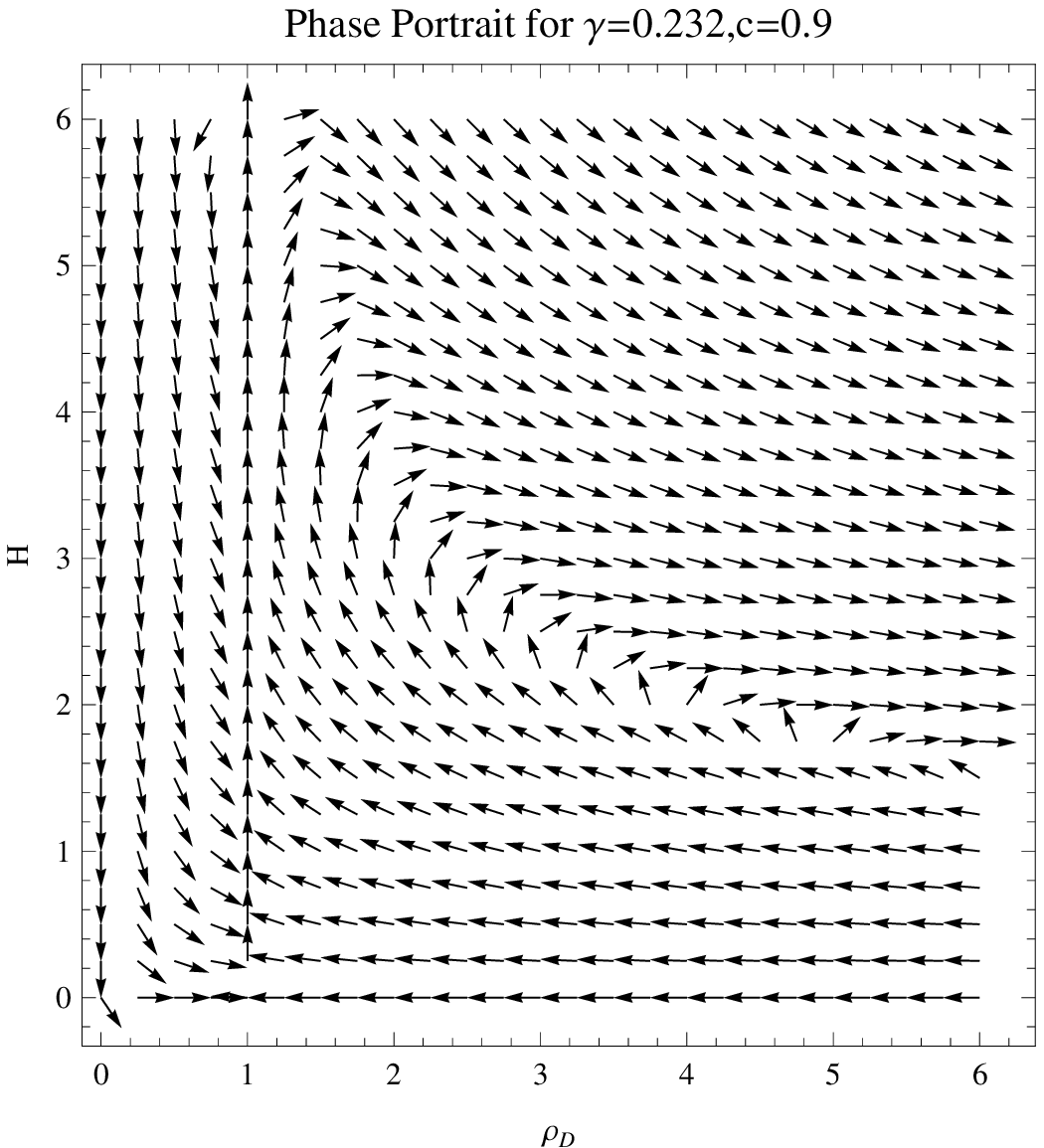}~~
\includegraphics[height=2.6in, width=2.6in]{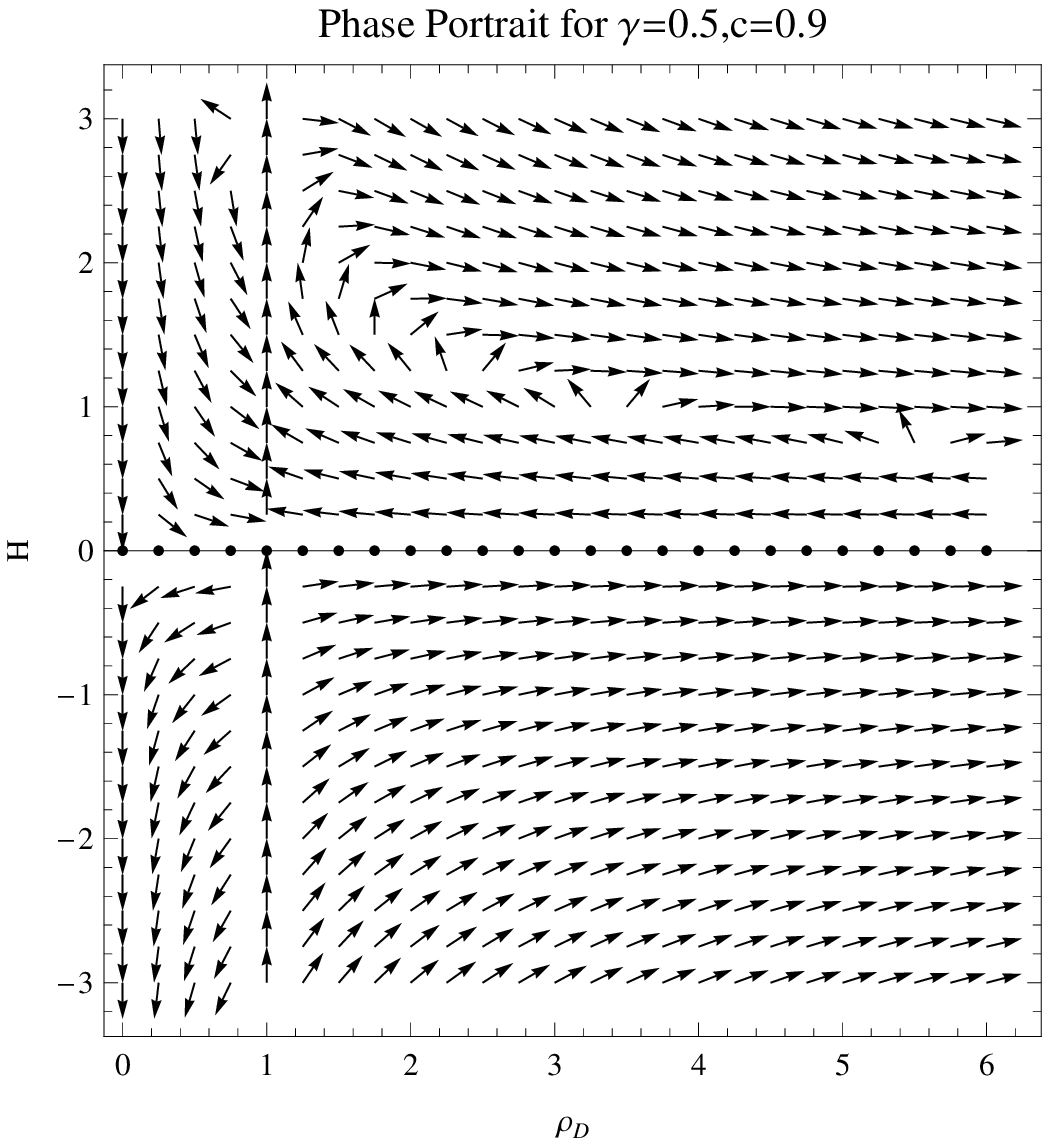}\\
\\
~~~~~~~~~~~~~~~~~~~~~~~~~~~~~~~~~~~~~~~~~~~~~~~~~~~~~~~~~~~~~~~~~Fig.2(a)
~~~~~~~~~~~~~~~~~~~~~~~~~~~~~~~~~~~~~~~~~~~~~~~~~~~~~~~~~~~~~~~~Fig.2(b)
\\\\

Fig.2(a)-2(b) represent the variation of $\Omega_D - H$. Here also
the negative coordinate of $H$ is not physically valid. But for
completeness of the system we have drawn the whole figure.

\hspace{1cm} \vspace{2cm}

\end{figure}

Hence equations (\ref{22}) and (\ref{24}) constitute an autonomous
system in the phase plane $\left(\Omega_{D}, ~H\right)$. The
possible critical points are\\
$(i)$ $H=0$, $\Omega_{D}$ is unrestricted,\\
$(ii)$ $\Omega_{D}=0,~H=0$,\\
$(iii)$ $\Omega_{D}=1,~H=0$ ,\\
$(iv)$$\Omega_{D}=1$, $H$ is unspecified and\\
$(v)$ $\Omega_{D}=c^{2},H=~\frac{1}{\gamma c^{2}}$.\\
The first three critical points correspond to static model of the
universe. In the first one the universe may have any amount of DE
while in the second one the universe does not have any DE. The
third and fourth critical points correspond to universe in phantom
era filled only with the DE. For the fifth critical point, the
ratio of the matter densities $u=\frac{1-c^{2}}{c^{2}}$ and
$\omega_{D}=-\frac{1}{c^{2}}<-1$. So the universe is again in the
phantom region. A detailed investigation of this critical point
will be done subsequently.

The conservation equations (\ref{5}) and (\ref{6}) can be written as
\begin{equation}\label{25}
\dot{\rho}_{m}=\rho_{m}\left[\gamma \rho_{D} - \sqrt{3\left(\rho_{m}+\rho_{D}\right)}\rho_{m}\right]
\end{equation}
\begin{equation}\label{26}
\dot{\rho_{D}}=-\rho_{D}\left[\gamma
\rho_{m}+\left(1+\omega_{D}\right)\sqrt{3\left(\rho_{m}+\rho_{D}\right)}\right]
\end{equation}
Apparently, we have similar situation as before,i.e., initially if
we assume to have sufficient DE then $\dot{\rho_D}<0$ and
$\dot{\rho_m}>0$ and subsequently $\dot{\rho_m}<0$. Note that
$\dot{\rho_m}=0$ along the curve
${\gamma}^2{\rho_D}^2-3{\rho_m}^3=3{\rho_D}{\rho_m}^2$. But due to
$\omega_{D}$ term in equation (\ref{26}) DE density begins to
increase in phantom era and we should have DE dominated universe
as expected in the present scenario. So both the energy densities
are of comparable magnitudes (i.e., $u\sim o(1)$) twice during the
evolution and give a possible explanation to the coincidence
problem.

To study the nature of the critical point
$\left(c^{2},~\frac{1}{\gamma c^{2}}\right)$ on the phase plane
$\left(\Omega_{D},~ H\right)$ we start with the linearized system
:
\begin{equation}\label{27}
\left.
\begin{array}{c}
\dot{x}=\frac{\left(1-c^{2}\right)}{c^{2}\gamma}\left(2x +3\gamma c^{4} y\right) \\\\
\dot{y}=\frac{\left(3-2c^{2}\right)x}{xc^{6}\gamma^{2}}+\frac{3\left(1-c^{2}\right)y}{2c^{2}\gamma}
\end{array}
\right\}
\end{equation}
where $x=\Omega_{D}-c^{2},~y=H-\frac{1}{\gamma c^{2}}$.

So the linearized matrix $A$ has
$$ tr(A)=\frac{7}{2}\frac{\left(1-c^{2}\right)}{c^{2}\gamma}>0$$ \\\\ $$det(A)=-\frac{3}{2}\frac{\left(1-c^{2}\right)}{c^{4}\gamma^{2}}<0$$

Thus for the linearized matrix $A$ both the eigen values are real
but of opposite sign and hence the critical point is of saddle
type and unstable in nature. The phase portrait for different
choices of $\gamma$ and $c^2$ are presented in figures 2(a) and
(b).

\section{Discussion}\label{chap5}

In the present work, we study the cosmological evolution of
interactive DM and DE in the background of a homogeneous and
isotropic FRW model of the universe. The DM is chosen in the form
of dust while for DE we choose holographic DE in the form of
perfect fluid having variable equation of state. The interaction
between DM and DE is chosen either as a linear combination of the
energy densities or in their product form. For both choices of the
interaction term, the evolution equations can be suitably
converted into an autonomous system and the critical points are
analyzed both analytically and graphically. Finally, the present
model of the universe shows partial solution of the coincidence
problem.

\end{document}